# Final State Resolved Quantum Predissociation Dynamics of $SO_2(\tilde{C}\,^1B_2)$ and Its Isotopomers via a Crossing with a Singlet Repulsive State

Changjian Xie,[1] Bin Jiang,[1,%] Jacek Kłos,[2] Praveen Kumar,[3] Millard H. Alexander,[2,4,*] Bill Poirier,[3,*] and Hua Guo[1,*]

[1]*Department of Chemistry and Chemical Biology, University of New Mexico, Albuquerque, New Mexico 87131, USA*

[2]*Department of Chemistry and Biochemistry, University of Maryland, College Park, MD 20742, USA*

[3]*Department of Chemistry and Biochemistry, Texas Tech University, Lubbock, TX 79409, USA*

[4]*Institute for Physical Science and Technology, University of Maryland, College Park, MD 20742, USA*

*%: present address: Department of Chemical Physics, University of Science and Technology of China, Hefei, Anhui, 230026 China*

**: corresponding authors, mha@umd.edu, 301 405 1823, bill.poirier@ttu.edu, 806 834 3099, hguo@unm.edu 505 277 1716.*

## Abstract

The fragmentation dynamics of predissociative $SO_2(\tilde{C}\,^1B_2)$ is investigated on an accurate adiabatic potential energy surface (PES) determined from high level ab initio data. This singlet PES features non-$C_{2v}$ equilibrium geometries for $SO_2$, which are separated from the SO($\tilde{X}\,^3\Sigma^-$) + O($^3P$) dissociation limit by a barrier resulting from a conical intersection with a repulsive singlet state. The ro-vibrational state distribution of the SO fragment is determined quantum mechanically for many predissociative states of several sulfur isotopomers of $SO_2$. Significant rotational and vibrational excitations are found in the SO fragment. It is shown that these fragment internal state distributions are strongly dependent on the predissociative vibronic states, and the excitation typically increases with the photon energy.

## I. Introduction

Produced by both nature and human activities, sulfur dioxide ($SO_2$) is an important species in Earth's atmosphere.[1] This molecule has also been found in the atmospheres of other planets and satellites in the solar system.[2-4] Its solar photochemistry has been linked to recently discovered sulfur mass-independent fractionation (S-MIF) in ancient rock samples,[5] which suggests the onset of the Great Oxygenation Event (GOE) in Earth's atmosphere.[5-7] The sudden change of the isotope ratios of sulfur in the rock record about 2500 million years ago could be due to the shielding of the atmospheric $SO_2$ from the solar ultra-violet (UV) rays by the nascent ozone layer. Despite extensive studies of the $SO_2$ photochemistry,[8-20] however, the chemical origin of the S-MIF in the rock records remains elusive. A comprehensive understanding of the photoabsorption and photoreaction of $SO_2$ as well as the subsequent chemical processes is clearly needed to gain insight into this important astrobiological event in the history of the Earth.

The strong UV absorption band of $SO_2$ in the 240-180 nm region is due to the dipole allowed $\tilde{C}\,^1B_2 \leftarrow \tilde{X}\,^1A_1$ transition.[21-23] The spectroscopy of this so-called *C* band has been extensively studied in the past, and a plethora of spectroscopic data exist above the band origin at 42573 cm$^{-1}$.[24-27] Above 218 nm, $SO_2$ is known to predissociate, as evidenced by a sudden decrease of fluorescence and broadening of the spectral features.[21, 24, 28-33] The dissociation dynamics to the SO($\tilde{X}\,^3\Sigma^-$) and O($^3P$) species has been studied by a number of groups, using different experimental techniques.[34-44] There is experimental evidence that the SO internal excitation varies with the excitation photon energy.[41] At 193 nm, the SO vibrational state distribution has been found to be bimodal, with a peak near $v$=2 with hot rotational distributions and another peak at $v$=5 with less rotational excitation.[43, 44]

Since the $\tilde{C}^1B_2$ state correlates diabatically to SO($\tilde{a}^1\Delta$) + O($^1D$), as shown in Figure 1, the production of the lower energy products SO($\tilde{X}^3\Sigma^-$) and O($^3P$) has to be induced by non-adiabatic crossings. One possibility is internal conversion to the $\tilde{X}^1A_1$ state, which then dissociates to SO($\tilde{X}^3\Sigma^-$) and O($^3P$) products. However, the location and characteristics of this conical intersection are yet to be defined, although an avoided crossing can be clearly seen from our recent work.[45] Alternatively, a conical intersection formed with the $\tilde{D}^1A'$ state, which is repulsive and correlates with the SO($\tilde{X}^3\Sigma^-$) + O($^3P$) asymptote, facilitates predissociation of $SO_2$($\tilde{C}^1B_2$).[31] Another non-adiabatic pathway is through intersystem crossing to the repulsive $2^3A'$ state, which also correlates to the same triplet asymptote.[31] As shown in the same figure, the former has a higher energy crossing. The involvement of the repulsive triplet state is supported by spin polarization experiments.[35, 46] On the other hand, evidence for the involvement of the single repulsive state in dissociation also exists.[33] In addition, the dissociation has also been attributed to the ground electronic state after internal conversion.[29, 32] The precise mechanism of dissociation is still unclear, and may be dependent on the photon wavelength.

While many theoretical studies have been carried out on the spectroscopic properties of $SO_2$ in the *C* band,[31, 45, 47-55] there has been no study of the dissociation dynamics. Such a process could ultimately yield the details relevant to the S-MIF effect. Indeed, the spectral range of 180-220 nm where a large S-MIF effect was found[8] coincides with the predissociation region in the *C* band, strongly suggesting the involvement of $SO_2$ photolysis. In this publication, we present the first study of the photodissociation dynamics of $SO_2$($\tilde{C}^1B_2$) following the crossing with a singlet repulsive state. Using an adiabatic model, the fragmentation dynamics of the predissociative states of $SO_2$ and several of its sulfur isotopomers just above the dissociation barriers is characterized by propagating wave packet on a recently developed ab initio PES.[45] The propagation was carried out to the asymptotic region to resolve the fragment internal state distribution. Although this model does not include all possible dissociation channels, it sheds valuable light on the dissociation dynamics and product energy disposal. This publication is organized as follows. The wave packet method is outlined in the next Section. The results are presented and discussed in Sec. III. Final conclusions are given in Sec. IV.

## II. Theory

In this work, we used the Chebyshev real wave packet method[56] to investigate the photodissociation dynamics of $SO_2(\tilde{C}^1B_2)$. The Hamiltonian in the SO + O′ Jacobi coordinates ($R$, $r$, $\gamma$) can be written as ($\hbar = 1$ hereafter),

$$\hat{H} = -\frac{1}{2\mu_R}\frac{\partial^2}{\partial R^2} - \frac{1}{2\mu_r}\frac{\partial^2}{\partial r^2} + \frac{\hat{j}^2}{2\mu_r r^2} + \frac{(\hat{J}-\hat{j})^2}{2\mu_R R^2} + V(R,r,\gamma)\,, \qquad (1)$$

where $R$ denotes the distance between the O′ atom and the center of mass of SO, $r$ is the S-O bond distance, and $\mu_R$ and $\mu_r$ are the corresponding reduced masses. $\hat{j}$ is the rotational angular momentum operator of SO and $\hat{J}$ ($J$=0 in the calculations) is the total angular momentum operator of $SO_2$. The spin-rotation coupling in $^3\Sigma$ state of SO is neglected and electron spin is considered as a spectator. $V(R, r, \gamma)$ is the adiabatic potential energy surface reported in our earlier work.[45]

The $\tilde{C}^1B_2 \leftarrow \tilde{X}^1A_1$ photoexcitation was simulated within the Condon approximation, in which the initial wave packet is given by the ground vibrational eigenfunction on the $\tilde{X}^1A_1$ state placed vertically on the $\tilde{C}^1B_2$ state PES. The initial wave packet is then propagated using the Chebyshev recursion relation:[56]

$$\Psi_k = 2D\hat{H}_s\Psi_{k-1} - D^2\Psi_{k-2}, \quad k \geq 2, \tag{2}$$

with $\Psi_1 = D\hat{H}_s\Psi_0$ and $\Psi_0$ is the initial wave packet. The Hamiltonian in Eq. (1) is scaled to the spectral range of (-1,1) by $\hat{H}_s = (\hat{H} - H^+)/H^-$, in which the spectral medium and half width ( $H^{\pm} = (H_{\max} \pm H_{\min})/2$ )were determined by the spectral extrema, $H_{\max}$ and $H_{\min}$, which can be readily estimated. The damping functions (*D*) were used to avoid reflection at the edge of the radial grids and its form is given below.

The absorption spectra were obtained from the discrete cosine Fourier transform of the Chebyshev autocorrelation function $C_k \equiv \langle \Psi_0 | \Psi_k \rangle$,[57]

$$S(E) = \frac{1}{\pi H^- \sin\vartheta} \sum_{k=0} (2 - \delta_{k,0}) \cos(k\vartheta) C_k, \tag{3}$$

where $E$ is the total energy and $\vartheta = \arccos E$ is the Chebyshev angle. A window function $e^{-\alpha^2 k^2/2}$ with $\alpha$=0.00002 was multiplied on Eq. (3) to smooth the absorption spectra.[53]

The distributions of all energetically available ro-vibrational products were obtained from discrete cosine Fourier transformations:[57]

$$A_f(E) = \left| \frac{1}{2\pi H^- \sin\vartheta} \sum_{k=0}^{N} (2 - \delta_{k0}) e^{-ik\vartheta} C_k^f \right|^2, \tag{4}$$

in which $C_k^f$ are the corresponding Chebyshev cross-correlation functions, which was calculated as

$$C_k^f = \left\langle \varphi_f \middle| \delta(R - R^{\infty}) \middle| \Psi_k \right\rangle, \tag{5}$$

where $\varphi_f$ are ro-vibrational eigenstates of the product SO, and $R$ was fixed at $R^{\infty}$=16.0 bohr in the product channel O + SO. The numerical parameters used in the calculations, including the form of the damping function, are summarized in Table 1.

## III. Results and discussion

Figure 2 shows the calculated absorption spectra of $^{32}SO_2$, $^{33}SO_2$, $^{34}SO_2$, and $^{36}SO_2$ from the vibrational ground state of the electronic ground state. It can be seen that there are strong oscillatory structures in the spectra corresponding to vibronic states of $SO_2(\tilde{C}\,^1B_2)$. The low-energy features of these spectra have been extensively discussed and assigned in our recent publications.[54, 55] There, the agreement with available experimental data is excellent, providing convincing evidence in support of the accuracy of the PES. In the higher energy range (>~53000 $cm^{-1}$), the widths of the resonances become broader, due apparently to the predissociation barrier resulted from a conical intersection with the repulsive $\tilde{D}^1A'$ state. The isotope-induced shifts of the absorption peaks have been extensively examined as it has been widely considered as the origin of the S-MIF effect.[11, 12, 18, 19, 52] As discussed in our recent work, the shifts of the spectra follow simple linear relationship with the photon energy,[55] in excellent agreement with known experimental data.[12, 19]

The adiabatic dissociation barrier from $SO_2(\tilde{C}\,^1B_2)$ to $SO(\tilde{X}\,^3\Sigma^-)$ + $O(^3P)$ is located at $R$=4.276 bohr, $r$=2.790 bohr, and $\gamma$=103.5° and has the energy of 53883 $cm^{-1}$, which is 10252 $cm^{-}$

$^{1}$ higher than the $SO_2(\tilde{C}^1B_2)$ minimum ($R$=3.456 bohr, $r$=2.828 bohr, and $\gamma$=119.6°). The barrier position is indicated by an arrow in the figure. It should be noted that the PES in the asymptotic region was modified to remove an artificial anisotropy in the asymptotic region near the SOO collinear geometry. This modification makes sure that the PES is isotropic in the asymptote. It has no effect on the vibrational distribution of the SO fragment, but shifts the rotational state distribution by no more than 5 rotational states.

The wave packet method used in this work allows the determination of the product state distribution in all required energies from a single wave packet propagation. To highlight the energy dependence, we have computed all product state distributions with an energy interval of 0.001 eV above the predissociation barrier. In Figures 3-6, the high-energy wing of the absorption spectra and the fraction of the product $SO(\tilde{X}^3\Sigma^-)$ vibrational populations in the photon energy range (53000-57600 cm$^{-1}$) are shown for $^{32}SO_2$, $^{33}SO_2$, $^{34}SO_2$, and $^{36}SO_2$. From the figures, it is clear that the widths of the resonance peaks are finite, and they gradually increase with energy, indicating shorter lifetimes. Unfortunately, the lifetimes of these resonance states cannot be accurately determined because they are often overlapping with each other. The SO vibrational state distribution is dependent on the individual resonances. However, a general trend exists that the vibrational excitation is higher at higher excitation energies. In the energy range < ~54000 cm$^{-1}$, the ground state product of SO dominates. As the excitation energy increases (54500~56000 cm$^{-1}$), the first excited vibrational state ($v$=1) of the SO product has the largest population. In even higher energies, the dominated vibrational state $v$=2 can be found in some energy points.

In Figure 7, the vibrational state distributions of the $SO(\tilde{X}^3\Sigma^-)$ fragment from the dissociation of $^{32}SO_2$ are displayed for six photon energies ($E_{hv}$=53706, 54024, 54189, 54263, 55950, and 56095 cm$^{-1}$). At these six energies, the most populated state is $v$=0, 1, or 2, respectively, much less than the energetically allowed maximal vibrational state indicated in the figure by an arrow. The rotational state distributions of the SO( $\tilde{X}\,^3\Sigma^-$) fragment are shown in Figure 8 at the same energies in the most populated vibrational state. These distributions suggest strong rotational excitation with peaks near *j*=~40.

The vibrational state excitation of the SO( $\tilde{X}\,^3\Sigma^-$) fragment at a particular energy is largely determined by the corresponding predissociation wavefunction at the dissociation barrier. Figure 9 (left panel) shows the wavefunctions cut in the (*R*, *r*) coordinates at the barrier geometry for $E_{hv}$=53706, 54189, and 56095 cm$^{-1}$. It can be seen from the figures that these wavefunctions are quite irregular inside the well, which indicate that they are difficult to assign presumably due to chaos. Such irregularity is already apparent at quite low energies. In addition, the wavefunctions at three energies possess different nodal structures (none, one, and two nodes for $E_{hv}$=53706, 54189, and 56095 cm$^{-1}$, respectively) along *r* direction in the exit channel, which are consistent with the dominated vibrational state in the distributions shown in Figure 7. The Franck-Condon overlaps between the wavefunctions in the *r* coordinate at the dissociation barrier and the SO vibrational eigenstates have been computed and they are displayed along with the calculated vibrational state distributions in Figure 7. As can be seen, these Franck-Condon overlaps predict the vibrational excitation in the SO( $\tilde{X}\,^3\Sigma^-$) product quite well. This indicates that the final state interaction in the dissociation channel is negligible and the energy release in the dissociation is largely in the translation and rotational (see below) degrees of freedom. This can be readily understood as the S-O bond length at the dissociation barrier is 2.790 bohr, very close to that of the SO fragment (2.809 bohr).

Figure 9 (right panel) shows the wavefunctions cut in the (*R*, $\gamma$) coordinates at the barrier geometry for $E_{hv}$=53706, 54189, and 56095 cm$^{-1}$. It can be seen that the wavefunctions are widely scattered along the $\gamma$ coordinates in the exit channel, which suggests the high excitations in the rotational mode of product SO. The high rotational excitation of the SO( $\tilde{X}^3\Sigma^-$ ) fragment originates from two sources. One is the bending energy of the $SO_2$ wavefunction at the dissociation barrier. For this, the projection of the wavefunction at the barrier onto the SO rotational state is computed at the energies and displayed in Figure 8 along with the fragment rotational state distributions. As shown, the energy in the $SO_2$ bending mode at the dissociation barrier is significantly smaller than that in the final state distribution. This is because the breaking of the S-O′ bond at the barrier, which is bent, exerts a torque of the departing SO, which channels energy into the fragment rotation during the final descend to the asymptote. In this case, the final state interaction is quite strong in the dissociation channel.

The internal excitation of the SO fragment can have a significant impact on its reactions with other atmospheric species, such as $H_2S$, OH and $O_2$. In a recent study, the reaction pathway for the SO + $O_2$ → $SO_3$ reaction has been explored.[18] It was shown that this reaction has a small thermal rate coefficient because of a number of barriers and bottlenecks along the reaction pathway. In another possible reaction, the SO can be reduced by $H_2S$, leading eventually to elemental sulfur ($S_8$).[58] It is conceivable that ro-vibrationally excited SO from the dissociation of $SO_2$ could significantly increase the reaction rate of these reactions. Such reactions are thought to propagate the S-MIF generated in the photoexcitation of $SO_2$.[18]

At this point, it is still difficult to compare our results with photodissociation experiments, as most measurements have been carried out at 193 nm or longer wavelengths.[34-44] As mentioned earlier, the barrier in our adiabatic PES is 53883 cm$^{-1}$, which corresponds to 191 nm of the photon wavelength. Consequently, this singlet channel discussed in this work is inaccessible in the previous experiments and the lower-energy triplet pathway should dominate. The triplet pathway cannot yet be investigated because the corresponding PES of the $2^3A'$ state and the spin-orbit couplings are not available. The dissociation via the singlet state barrier is shown above to produce vibrationally unexcited SO just above the dissociation barrier, which is inconsistent with the experimental observations that the *v*=2 state dominates. Hence, our results support, albeit indirectly, that the dissociation near 193 nm is due to intersystem crossing via a triplet repulsive state.

## IV. Conclusions

In this work, we explore the dissociation dynamics of several isotopomers of $SO_2$ following the excitation to the $\tilde{C}^1B_2$ state. The predissociation via a barrier formed by a conical intersection between the $\tilde{C}^1B_2$ state and a repulsive singlet ($\tilde{D}^1A'$) state is investigated. This system provides an ideal proving ground for studying state-selective unimolecular dissociation. Our results indicate that the SO($\tilde{X}^3\Sigma^-$) fragment internal state distribution depends strongly on the vibronic state it is initiated. At low excitation energies just above the dissociation barrier, the SO($\tilde{X}^3\Sigma^-$) fragment is dominated by the ground vibrational state. However, as energy increases, more excited vibrational states are populated. The fragment vibrational state distributions are largely determined by the $SO_2$ vibronic wavefunction at the dissociation barrier, suggesting small final state interaction. On the other hand, the SO rotation is highly excited, due to contributions from the bending wavefunction and a torque acting on the SO fragment as the S-O' bond is cleaved.

The dynamics discussed in this work should only be considered as the first step towards a full understanding of the predissociation dynamics of $SO_2$($\tilde{C}^1B_2$). To that end, the other two dissociation channels, namely the internal conversion to the $\tilde{X}\,^1A_1$ state and the intersystem crossing via a triplet ($2\,^3A'$) repulsive state, have to be included in the model. The former requires the location and characterization of the conical intersection seam between the two electronic states. The latter demands the mapping of the $2\,^3A'$ state PES and the spin-orbit coupling that facilitates the intersystem crossing. Work in this direction has already started in our groups.

**Acknowledgements**: This work was supported by a research grant (NNX13AJ49G-EXO) from NASA Astrobiology. H.G. also thanks the U.S. Department of Energy for partial support (DE-SC0015997).

Table 1. Numerical parameters used in the wave packet calculations. Atomic units are used unless stated otherwise.

| |
|---|
| Grid/basis range and size:<br>$R \in [1.0, 20.0]$, $N_R = 345$<br>$r \in [2.0, 6.5]$, $N_r = 85$<br>$j_{\max} = 119$, $N_j = 120$ |
| Damping along $R$:<br>$D_R = \exp\left[-0.005(R-16.5)^2\right]$, for $R > 16.5$<br>$= 1$, otherwise<br>Damping along $r$:<br>$D_r = \exp\left[-0.009(r-4.5)^2\right]$, for $r > 4.5$<br>$= 1$, otherwise |
| Propagation step: 250000 |

References:

1. Wayne, R. P. *Chemistry of Atmospheres*. Oxford University Press: Oxford, 2000.
2. Kumar, S. Photochemistry of $SO_2$ in the atmosphere of Io and implications on atmospheric escape, *J. Geophys. Res.-Space Phys.* **1982,** *87*, 1677-1684.
3. Halevy, I.; Zuber, M. T.; Schrag, D. P. A sulfur dioxide climate feedback on early Mars, *Science* **2007,** *318*, 1903-1907.
4. Zhang, X.; Liang, M.-C.; Montmessin, F.; Bertaux, J.-L.; Parkinson, C.; Yung, Y. L. Photolysis of sulphuric acid as the source of sulphur oxides in the mesosphere of Venus, *Nat. Geosci.* **2010,** *3*, 834-837.
5. Farquhar, J.; Bao, H.; Thiemens, M. H. Atmospheric influence of Earth's earliest sulfur cycle, *Science* **2000,** *289*, 756-758.
6. Pavlov, A. A.; Kasting, J. F. Mass-independent fractionation of sulfur isotopes in archean sediments: Strong evidence ofr an anoxic Archean atmosphere, *Astrobio.* **2002,** *2*, 27-41.
7. Halevy, B.; Johnston, D. T.; Schrag, D. P. Explaining the structure of the Archean mass-independent sulfur isotope record, *Science* **2010,** *329*, 204-207.
8. Farquhar, J.; Savarino, J.; Airieau, S.; Thiemens, M. H. Observation of wavelength-sensitive mass-independent sulfur isotope effects during $SO_2$ photolysis: Implications for the early atmosphere, *J. Geophys. Res.* **2001,** *106*, 32829-32839.
9. Farquhar, J.; Peters, M.; Johnston, D. T.; Strauss, H.; Masterson, A.; Wiechert, U.; Kaufman, A. J. Isotopic evidence for Mesoarchaean anoxia and changing atmospheric sulphur chemistry, *Nature* **2007,** *449*, 706-709.
10. Lyons, J. R. Mass-independent fractionation of sulfur isotopes by isotope-selective photodissociation of $SO_2$, *Geophys. Res. Lett.* **2007,** *34*, L22811.
11. Lyons, J. R. Photolysis of long-lived predissociative molecules as a source of mass-independent isotope fractionation, *Adv. Quantum Chem.* **2008,** *55*, 57–74.
12. Danielache, S. O.; Eskebjerg, C.; Johnson, M. S.; Ueno, Y.; Yoshida, N. High-precision spectroscopy of $^{32}S$, $^{33}S$, and $^{34}S$ sulfur dioxide: Ultraviolet absorption cross sections and isotope effects, *J. Geophys. Res.* **2008,** *113*, D17314.
13. Masterson, A. L.; Farquhar, J.; Wing, B. A. Sulfur mass-independent fractionation patterns in the broadband UV photolysis of sulfur dioxide: Pressure and third body effects, *Earth Plant. Sci. Lett.* **2011,** *306*, 253-260.
14. Danielache, S. O.; Hattori, S.; Johnson, M. S.; Ueno, Y.; Nanbu, S.; Yoshida, N. Photoabsorption cross-section measurements of $^{32}S$, $^{33}S$, $^{34}S$, and $^{36}S$ sulfur dioxide for the $B^1B_1$-$X^1A_1$ absorption band, *J. Geophys. Res.-Atmos.* **2012,** *117*, D24301.
15. Ono, S.; Whitehill, A. R.; Lyons, J. R. Contribution of isotopologue self-shielding to sulfur mass-independent fractionation during sulfur dioxide photolysis, *J. Geophys. Res.-Atmos.* **2013,** *118*, 2444-2454.
16. Whitehill, A. R.; Xie, C.; Hu, X.; Xie, D.; Guo, H.; Ono, S. Vibronic origin of mass-independent isotope effect in photoexcitation of $SO_2$, and the implications to the early Earth's atmosphere, *Proc. Natl. Acad. Sci. USA* **2013,** *110*, 17697–17702.
17. Hattori, S.; Schmidt, J. A.; Johnson, M. S.; Danielache, S. O.; Yamada, A.; Ueno, Y.; Yoshida, N. $SO_2$ photoexcitation mechanism links mass-independent sulfur isotopic fractionation in cryospheric sulfate to climate impacting volcanism, *Proc. Natl. Acad. Sci. USA* **2013,** *110*, 17656-17661.
18. Whitehill, A. R.; Jiang, B.; Guo, H.; Ono, S. $SO_2$ photolysis as a source for sulfur mass-independent isotope signatures in stratospehric aerosols, *Atmos. Chem. Phys.* **2015,** *15*, 1843–1864.

19. Endo, Y.; Danielache, S. O.; Ueno, Y.; Hattori, S.; Johnson, M. S.; Yoshida, N.; Kjaergaard, H. G. Photoabsorption cross-section measurements of $^{32}S$, $^{33}S$, $^{34}S$, and $^{36}S$ sulfur dioxide from 190 to 220 nm, *J. Geophys. Res.-Atmos.* **2015,** *120*, 2546-2557.
20. Endo, Y.; Ueno, Y.; Aoyama, S.; Danielache, S. O. Sulfur isotope fractionation by broadband UV radiation to optically thin $SO_2$ under reducing atmosphere, *Earth Planet. Sci. Lett.* **2016,** *453*, 9-22.
21. Okabe, H. Fluorescence and predissociation of sulfur dioxide, *J. Am. Chem. Soc.* **1971,** *93*, 7095-7096.
22. Brand, J. C. D.; Chiu, P. H.; Hoy, A. R. Sulfur dioxide: rotational constants and asymmetric structure of the $C^1B_2$ state, *J. Mol. Spectrosc.* **1976,** *60*, 43-56.
23. Hoy, A. R.; Brand, J. C. D. Asymmetric structure and force field of the $^1B_2(1A')$ state of sulphur dioxide, *Mol. Phys.* **1978,** *36*, 1409-1420.
24. Yamanouchi, K.; Okunishi, M.; Endo, Y.; Tsuchiya, S. Laser induced fluorescence spectroscopy of the C-X band of jet coold $SO_2$. Rotational and vibrational analysis in the 235-210 nm region, *J. Mol. Struct.* **1995,** *352/353*, 541-559.
25. Park, G. B.; Womack, C. C.; Whitehill, A. R.; Jiang, J.; Ono, S.; Field, R. W. Millimeter-wave optical double resonance schemes for rapid assignment of perturbed spectra, with applications to the $\tilde{C}^1B_2$ state of $SO_2$, *J. Chem. Phys.* **2015,** *142*, 144201.
26. Park, G. B.; Jiang, J.; Saladrigas, C. A.; Field, R. W. Observation of $b_2$ symmetry vibrational levels of the $SO_2$ $\tilde{C}^1B_2$ state: Vibrational level staggering, Coriolis interactions, and rotation-vibration constants, *J. Chem. Phys.* **2016,** *144*, 144311.
27. Jiang, J.; Park, G. B.; Field, R. W. The rotation-vibration structure of the $SO_2\tilde{C}^1B_2$ state explained by a new internal coordinate force field, *J. Chem. Phys.* **2016,** *144*, 144312.
28. Hui, M.-H.; Rice, S. A. Decay of flurorescence from single vibronic states of $SO_2$, *Chem. Phys. Lett.* **1972,** *17*, 474-478.
29. Ivanco, M.; Hager, J.; Sharfin, W.; Wallace, S. C. Quantum interference phenomena in the radiative decay of the $C(^1B_2)$ state of $SO_2$, *J. Chem. Phys.* **1983,** *78*, 6531-6540.
30. Ebata, T.; Nakazawa, O.; Ito, M. Rovibrational dependence of the predissociation in the $C^1B_2$ state of $SO_2$, *Chem. Phys. Lett.* **1988,** *143*, 31-37.
31. Katagiri, H.; Sako, T.; Hishikawa, A.; Yazaki, T.; Onda, K.; Yamanouchi, K.; Yoshino, K. Experimental and theoretical exploration of photodissociation of $SO_2$ via the $C^1B_2$ state: identification of the dissociation pathway, *J. Mol. Struc.* **1997,** *413/414*, 589-614.
32. Okazaki, A.; Ebata, T.; Mikami, N. Degenerate four-wave mixing and photofragment yield spectroscopic study of jet-cooled $SO_2$ in the $C^1B_2$ state: internal conversion followed by dissociation in the X state, *J. Chem. Phys.* **1997,** *107*, 8752-8758.
33. Ray, P. C.; Arendt, M. F.; Butler, L. J. Resonance emission spectroscopy of predissociating $SO_2$ $C(^1B_2)$: coupling with a repulsive $^1A_1$ state near 200 nm, *J. Chem. Phys.* **1998,** *109*, 5221-5223.
34. Freedman, A.; Yang, S.-C.; Bersohn, R. Translational energy distribution of the photofragments of $SO_2$ at 193 nm, *J. Chem. Phys.* **1979,** *70*, 5313-5314.
35. Kanamori, H.; Butler, J. E.; Kawaguchi, K.; Yamada, C.; Hirota, E. Spin polarization in SO photochemically generated from $SO_2$, *J. Chem. Phys.* **1985,** *83*, 611-615.
36. Kawasaki, M.; Sato, H. Photodissociation of molecular beams of $SO_2$ at 193 nm, *Chem. Phys. Lett.* **1987,** *139*, 585-588.
37. Felder, P.; Effenhauser, C. S.; Haas, B. M.; Huber, J. R. Photodissociation of sulfur dioxide at 193 nm, *Chem. Phys. Lett.* **1988,** *148*, 417-422.
38. Chen, X.; Asmar, F.; Wang, H.; Weiner, B. R. Nascent sulfur monoxide $(X^3\Sigma^-)$ vibrational distributions from the photodissociation of sulfur dioxide, sulfonyl chloride, and dimethylsulfoxide at 193 nm, *J. Phys. Chem.* **1991,** *95*, 6415-6417.

39. Becker, S.; Braatz, C.; Lindner, J.; Tiemann, E. State specific photodissociation of $SO_2$ and state selective detection of the SO fragment, *Chem. Phys. Lett.* **1993,** *208*, 15-20.
40. Braatz, C.; Tiemann, E. State-to-state dissociation of $SO_2$ in $C^1B_2$: rotational distributions of the fragment SO, *Chem. Phys.* **1998,** *229*, 93-105.
41. Cosofret, B. R.; Dylewski, S. M.; Houston, P. L. Changes in the vibrational population of $SO(^3\Sigma^-)$ from the photodissociation of $SO_2$ between 202 and 207 nm, *J. Phys. Chem. A* **2000,** *104*, 10240-10246.
42. Gong, Y.; Makarov, V. I.; Weiner, B. R. Time-resolved Fourier transform infrared study of the 193 nm photolysis of $SO_2$, *Chem. Phys. Lett.* **2003,** *378*, 493-502.
43. Yamasaki, K.; Taketani, F.; Sugiura, K.; Tokue, I.; Tsuchiya, K. Vibrational relaxation of $SO(X^3\Sigma^-$, v = 1−5) and nascent vibrational distributions of SO generated in the photolysis of $SO_2$ at 193.3 nm, *J. Phys. Chem. A* **2004,** *108*, 2382-2388.
44. Ma, J.; Wilhelm, M. J.; Smith, J. M.; Dai, H.-L. Photolysis (193 nm) of $SO_2$: Nascent product energy distribution examined through IR emission, *J. Phys. Chem. A* **2012,** *116*, 166-173.
45. Kłos, J.; Alexander, M. H.; Kumar, P.; Poirier, B.; Jiang, B.; Guo, H. New ab initio adiabatic potential energy surfaces and bound state calculations for the singlet ground $\tilde{X}^1A_1$ and excited $\tilde{C}^1B_2(2^1A')$ states of $SO_2$, *J. Chem. Phys.* **2016,** *144*, 174301.
46. Brouard, M.; Cireasa, R.; Clark, A. P.; Preston, T. J.; Vallance, C.; Groenenboom, G. C.; Vasyutinskii, O. S. $O(^3P_J)$ aligment from the photodissociation of $SO_2$ at 193 nm, *J. Phys. Chem. A* **2004,** *108*, 7965-7976.
47. Kamiya, K.; Matsui, H. Theoretical studies on the potential energy surfaces of $SO_2$: Electronic states for photodissociation from the $C^1B_2$ state, *Bull. Chem. Soc. Japan* **1991,** *64*, 2792-2801.
48. Nachtigall, P.; Hrusak, J.; Bludsky, O.; Iwata, S. Investigation of the potential energy surfaces for the ground $X^1A_1$ and excited $C^1B_2$ electronic states of $SO_2$, *Chem. Phys. Lett.* **1999,** *303*, 441-446.
49. Bludsky, O.; Nachtigall, P.; Hrusak, J.; Jensen, P. The calculation of the vibrational states of $SO_2$ in the $C^1B_2$ electronic state up to the SO+O dissociation limit, *Chem. Phys. Lett.* **2000,** *318*, 607-613.
50. Parsons, B.; Butler, L. J.; Xie, D.; Guo, H. A combined experimental and theoretical study of resonance emission spectra of $SO_2(C^1B_2)$, *Chem. Phys. Lett.* **2000,** *320*, 499-506.
51. Xie, D.; Guo, H.; Bludsky, O.; Nachtigall, P. Absorption and resonance emission spectra of $SO_2$(X/C) calculated from ab initio potential energy and transition dipole moment surfaces, *Chem. Phys. Lett.* **2000,** *329*, 503-510.
52. Ran, H.; Xie, D.; Guo, H. Theoretical studies of $C^1B_2$ absorption spectra of $SO_2$ isotopomers, *Chem. Phys. Lett.* **2007,** *439*, 280-283.
53. Tokue, I.; Nanbu, S. Theoretical studies of absorption cross sections for the $C\ ^1B_2$ - $X\ ^1A_1$ system of sulfur dioxide and isotope effects, *J. Chem. Phys.* **2010,** *132*, 024301.
54. Kumar, P.; Jiang, B.; Guo, H.; Kłos, J.; Alexander, M. H.; Poirier, B. Photoabsorption assignments for the $\tilde{C}^1B_2 \leftarrow \tilde{X}^1A_1$ vibronic transitions of $SO_2$, using new ab initio potential energy and transition dipole surfaces, *J. Phys. Chem. A* **2017,** *121*, 1012−1021.
55. Jiang, B.; Kumar, P.; Kłos, J.; Alexander, M. H.; Poirier, B.; Guo, H. First-principles C band absorption spectra of $SO_2$ and its isotopologues, *J. Chem. Phys.* **2017,** *146*, 154305.
56. Guo, H. Recursive solutions to large eigenproblems in molecular spectroscopy and reaction dynamics, *Rev. Comput. Chem.* **2007,** *25*, 285-347.
57. Guo, H. A time-independent theory of photodissociation based on polynomial propagation, *J. Chem. Phys.* **1998,** *108*, 2466-2472.
58. Kumar, M.; Francisco, J. S. Elemental sulfur aerosol-forming mechanism, *Proc. Natl. Acad. Sci. U. S. A.* **2017,** *114*, 864-869.

**Figure captions:**

Figure 1. Schematic potential energy curves for several low-lying electronic states of $SO_2$ along the dissociation coordinate. The singlet adiabatic potential used in this work is depicted by the red bold curve.

Figure 2. Absorption spectra of $^{32}SO_2$, $^{33}SO_2$, $^{34}SO_2$, and $^{36}SO_2$. The arrows denote the position of the barrier, which are 52351, 52359, 52366, and 52379 $cm^{-1}$ for $^{32}SO_2$, $^{33}SO_2$, $^{34}SO_2$, and $^{36}SO_2$, respectively.

Figure 3. Absorption spectrum and product vibrational state distributions in the predissociative energy range (53100~57500 $cm^{-1}$) of $^{32}SO_2$.

Figure 4. Absorption spectrum and product vibrational state distributions in the predissociative energy range (53100~57500 $cm^{-1}$) of $^{33}SO_2$.

Figure 5. Absorption spectrum and product vibrational state distributions in the predissociative energy range (53100~57500 $cm^{-1}$) of $^{34}SO_2$.

Figure 6. Absorption spectrum and product vibrational state distributions in the predissociative energy range (53100~57500 $cm^{-1}$) of $^{36}SO_2$.

Figure 7. SO vibrational state distributions (red lines) from the predissociation of $^{32}SO_2$ at $E_{hv}$=53706, 54024, 54189, 54263, 55950, and 56095 $cm^{-1}$. Blue dashed lines denote the Franck-Condon overlaps between the wavefunctions in the $r$ coordinate at the dissociation barrier and the SO vibrational eigenstates. The arrows denote the highest available vibrational states of SO.

Figure 8. SO rotational distributions (red lines) from the predissociation of $^{32}SO_2$ at $E_{hv}$=53706, 54024, 54189, 54263, 55950, and 56095 $cm^{-1}$. The vibrational states of SO at these energies are fixed at the most populated vibrational states as shown in Fig. 7. Blue dashed lines denote the Franck-Condon overlaps between the wavefunctions in the $\gamma$ coordinate at the dissociation barrier and the SO free rotor eigenstates.

Figure 9. Wavefunctions of $^{32}SO_2$ extracted at $E_{h\nu}$=53706, 54189, and 56095 cm$^{-1}$, (left panel) with $\gamma$ fixed at that of the barrier geometry structure (103.5˚); (right panel) with $r$ fixed at that of the barrier geometry structure (2.790 bohr).

Fig. 1

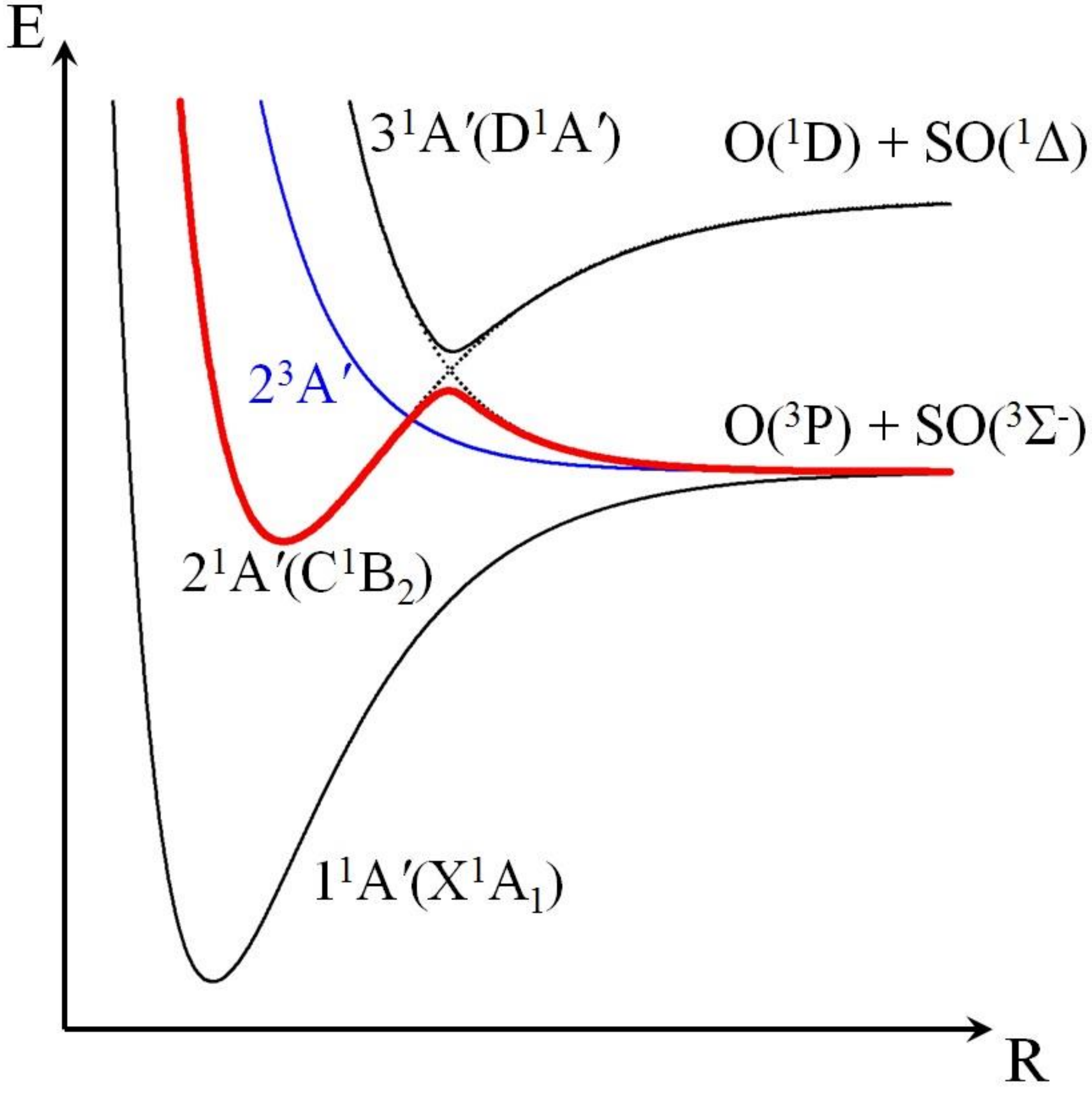

Fig. 2

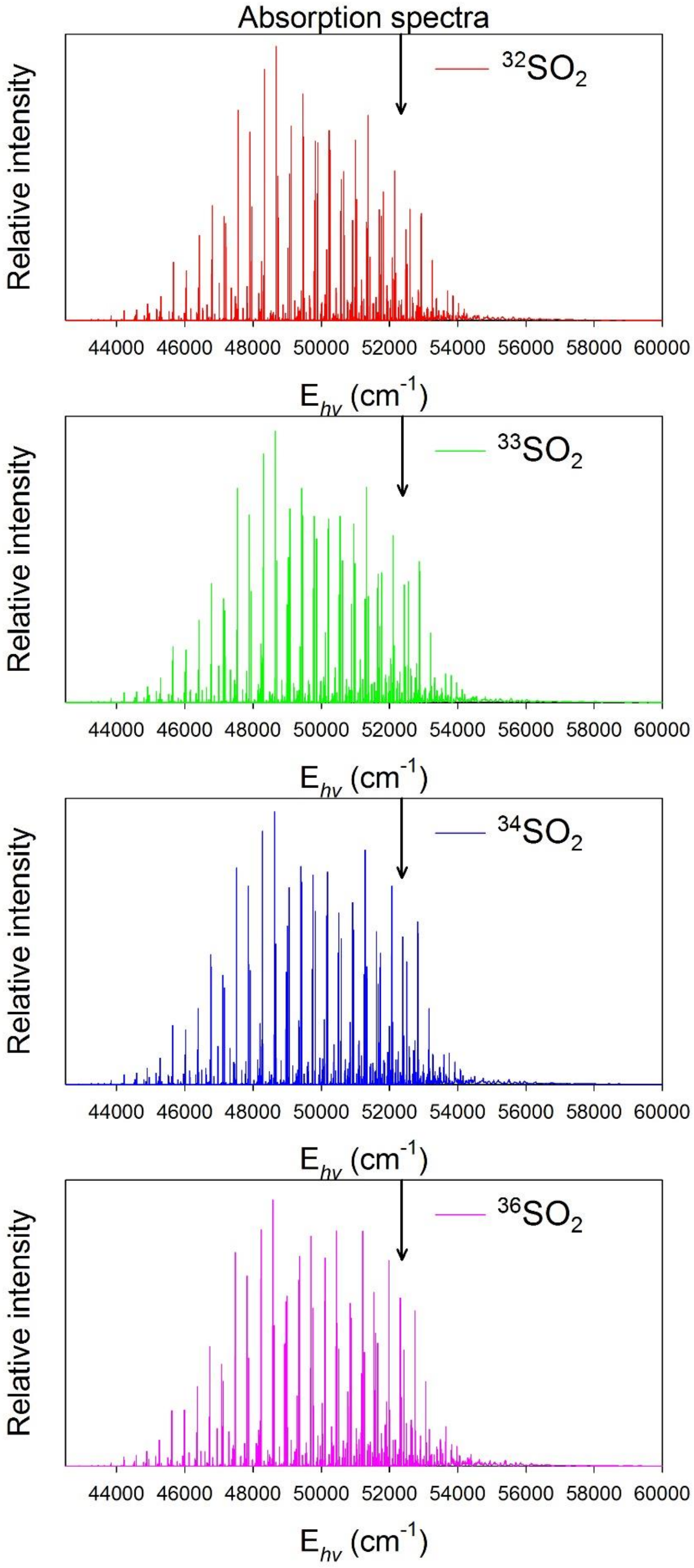

Absorption spectra
Relative intensity
$^{32}SO_2$
44000 46000 48000 50000 52000 54000 56000 58000 60000
$E_{h\nu}$ (cm$^{-1}$)
Relative intensity
$^{33}SO_2$
44000 46000 48000 50000 52000 54000 56000 58000 60000
$E_{h\nu}$ (cm$^{-1}$)
Relative intensity
$^{34}SO_2$
44000 46000 48000 50000 52000 54000 56000 58000 60000
$E_{h\nu}$ (cm$^{-1}$)
Relative intensity
$^{36}SO_2$
44000 46000 48000 50000 52000 54000 56000 58000 60000
$E_{h\nu}$ (cm$^{-1}$)

Fig. 3

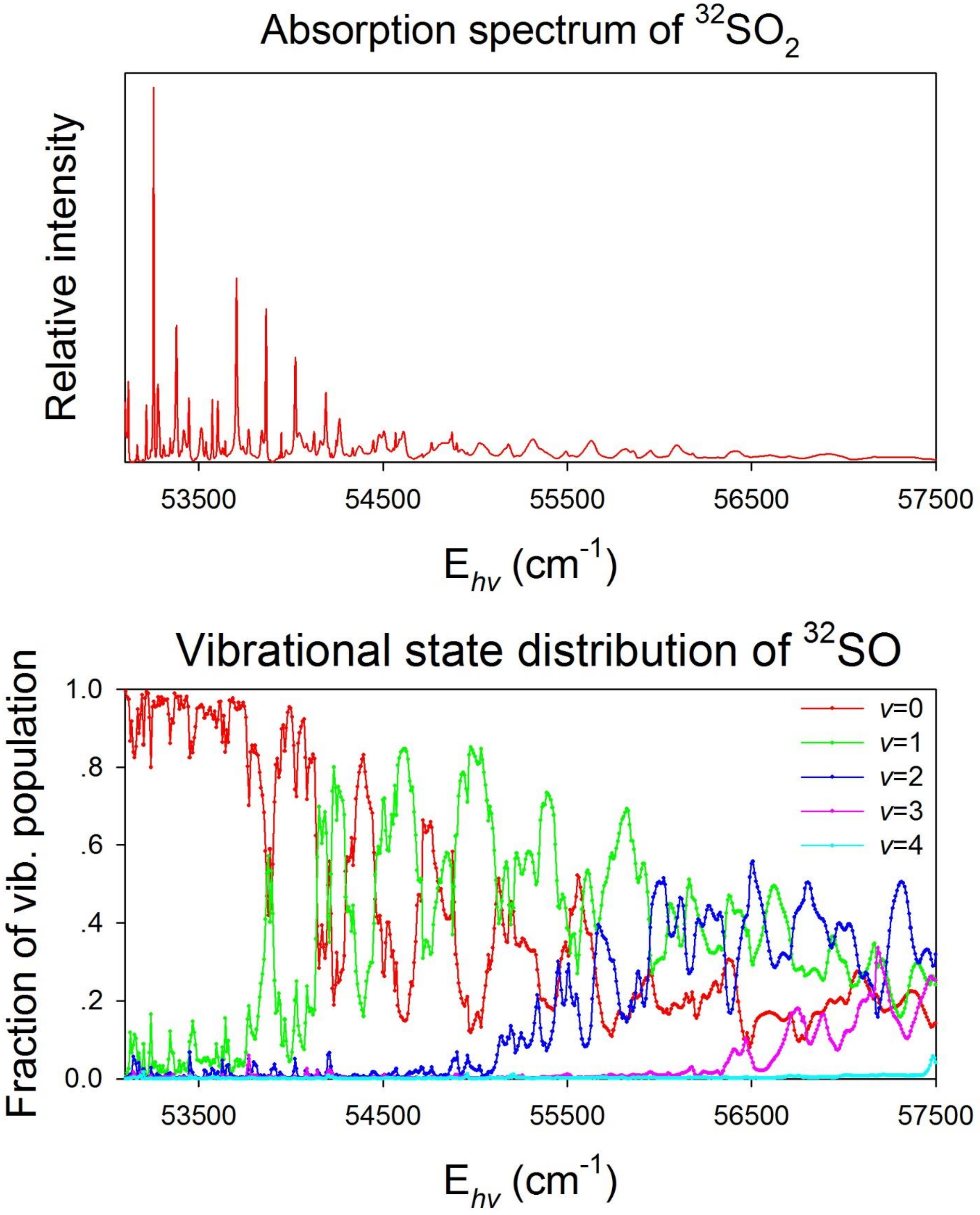

Fig. 4

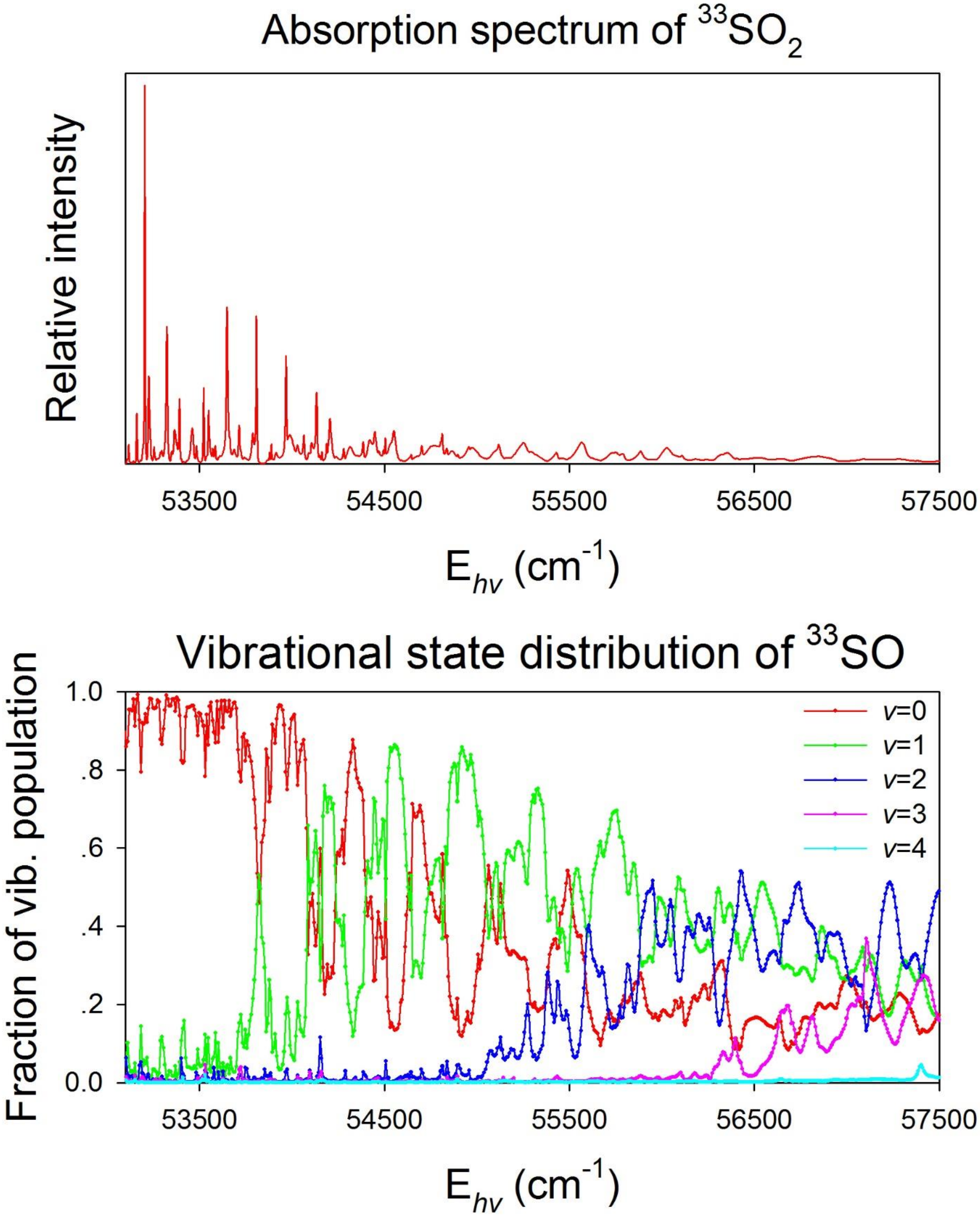

Fig.5

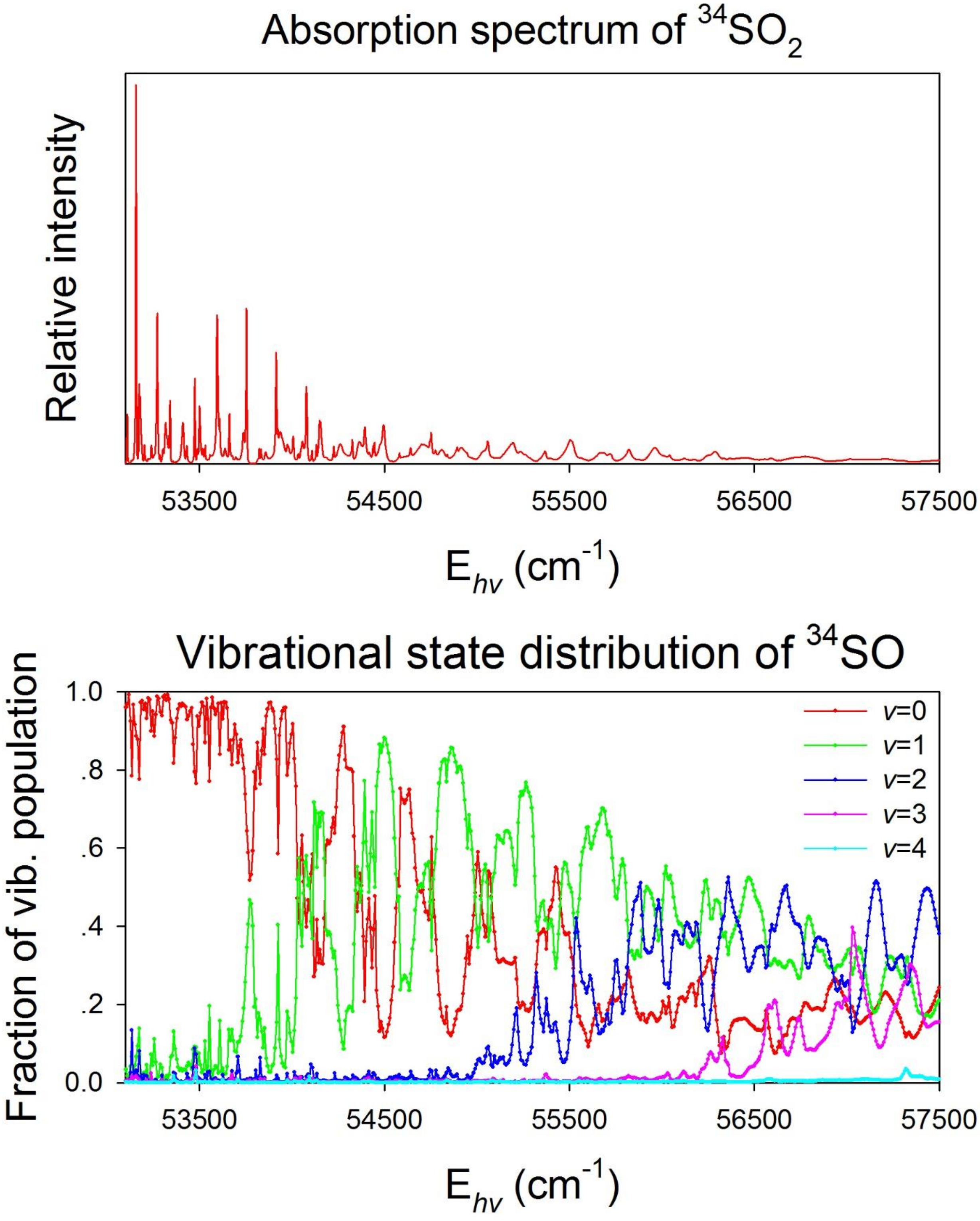

Fig. 6

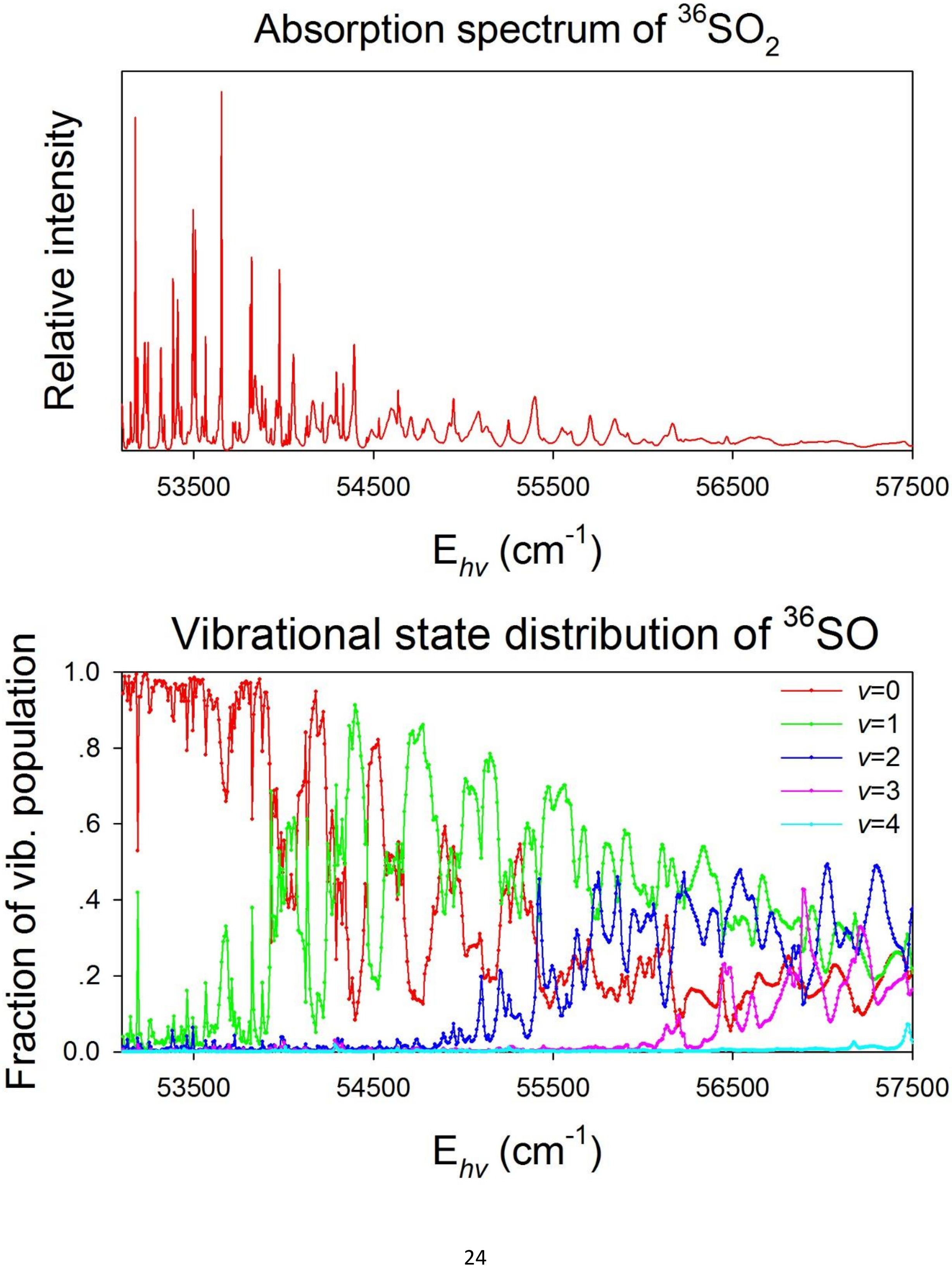

Fig. 7

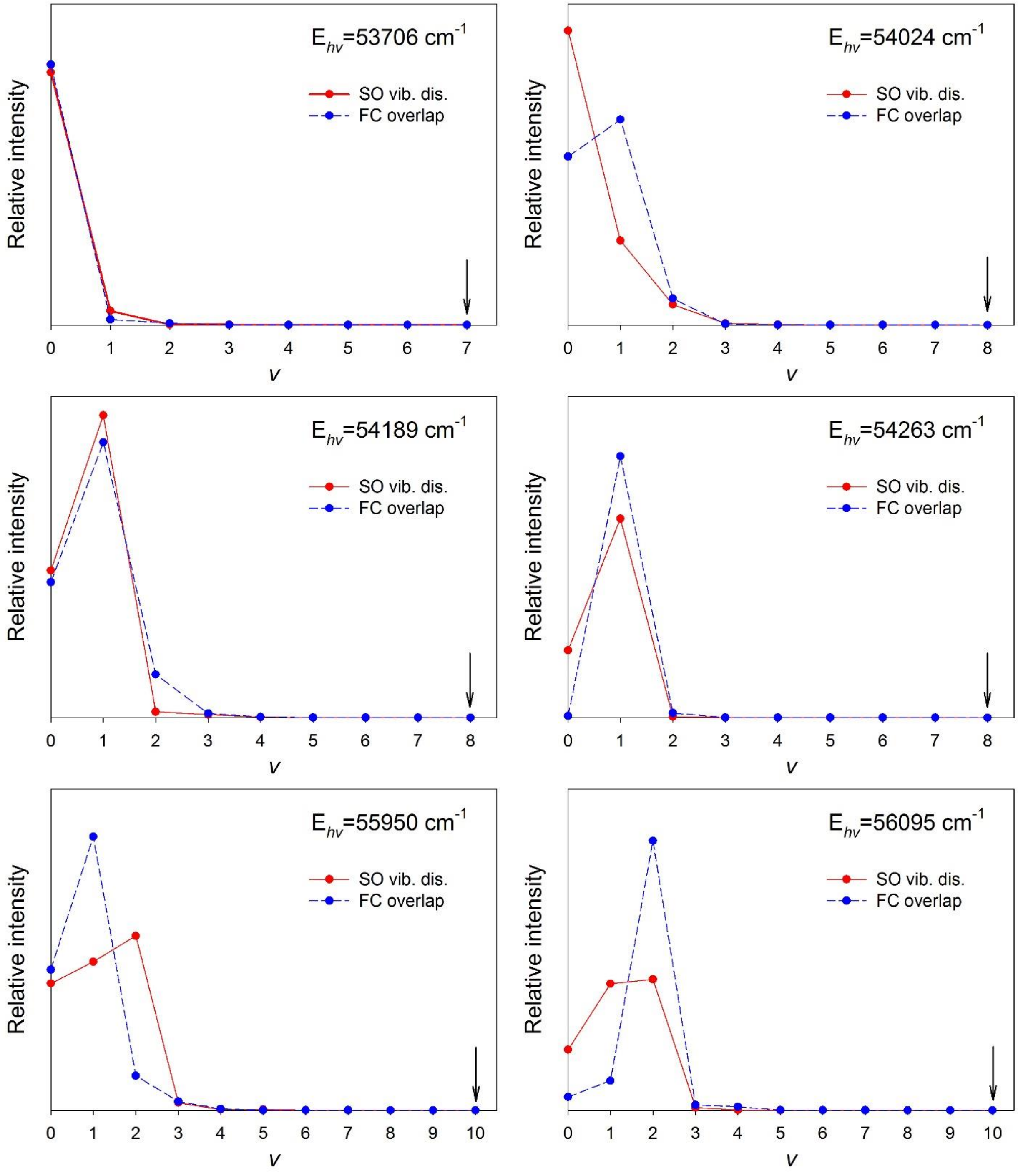

$E_{hv}$=53706 cm$^{-1}$
SO vib. dis.
FC overlap
Relative intensity
0 1 2 3 4 5 6 7
v
$E_{hv}$=54024 cm$^{-1}$
SO vib. dis.
FC overlap
Relative intensity
0 1 2 3 4 5 6 7 8
v
$E_{hv}$=54189 cm$^{-1}$
SO vib. dis.
FC overlap
Relative intensity
0 1 2 3 4 5 6 7 8
v
$E_{hv}$=54263 cm$^{-1}$
SO vib. dis.
FC overlap
Relative intensity
0 1 2 3 4 5 6 7 8
v
$E_{hv}$=55950 cm$^{-1}$
SO vib. dis.
FC overlap
Relative intensity
0 1 2 3 4 5 6 7 8 9 10
v
$E_{hv}$=56095 cm$^{-1}$
SO vib. dis.
FC overlap
Relative intensity
0 1 2 3 4 5 6 7 8 9 10
v

Fig. 8

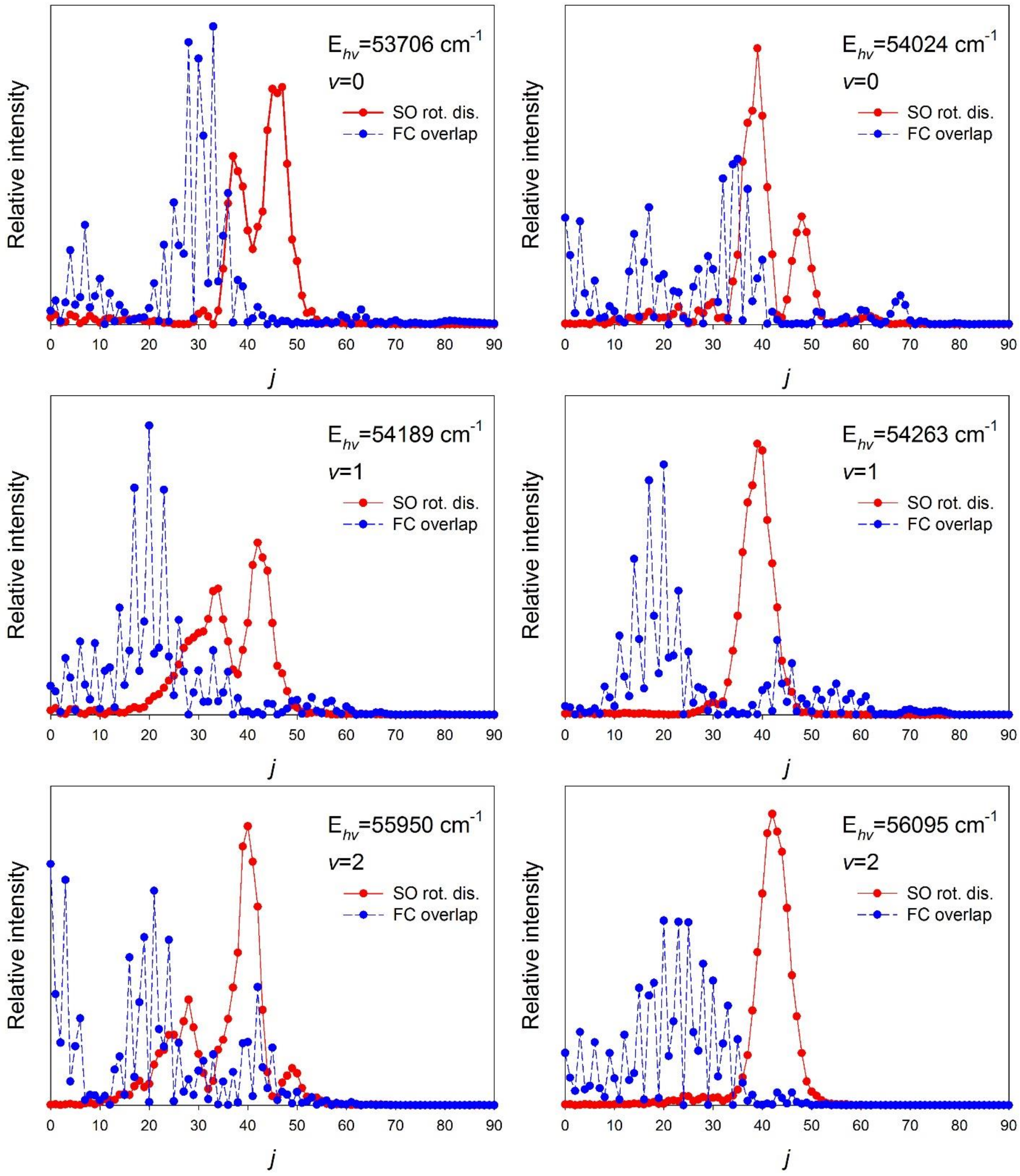

E$_{hv}$=53706 cm$^{-1}$
$v$=0
SO rot. dis.
FC overlap
E$_{hv}$=54024 cm$^{-1}$
$v$=0
SO rot. dis.
FC overlap
E$_{hv}$=54189 cm$^{-1}$
$v$=1
SO rot. dis.
FC overlap
E$_{hv}$=54263 cm$^{-1}$
$v$=1
SO rot. dis.
FC overlap
E$_{hv}$=55950 cm$^{-1}$
$v$=2
SO rot. dis.
FC overlap
E$_{hv}$=56095 cm$^{-1}$
$v$=2
SO rot. dis.
FC overlap
Relative intensity
$j$
0 10 20 30 40 50 60 70 80 90

Fig. 9

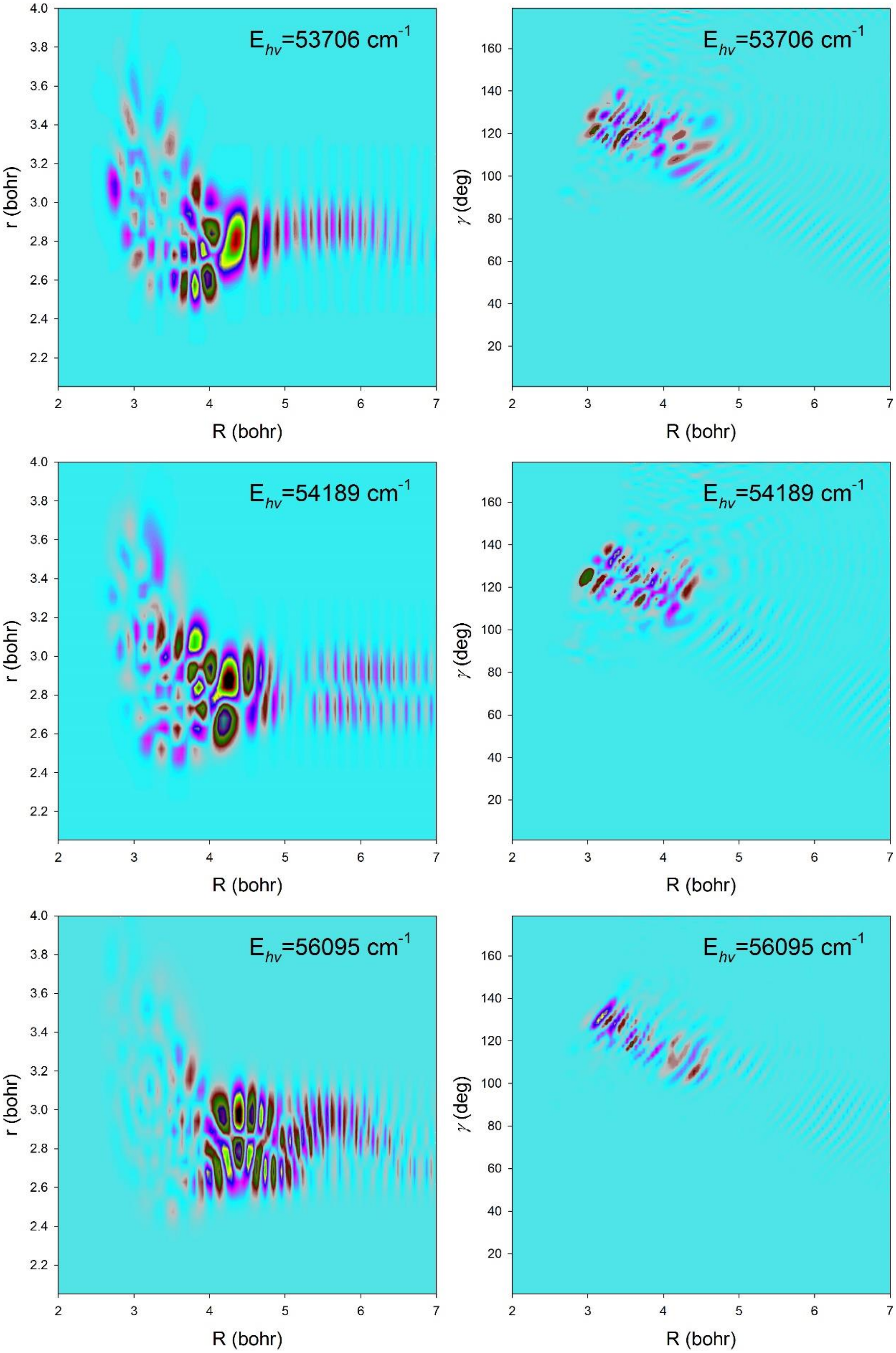

TOC graphic

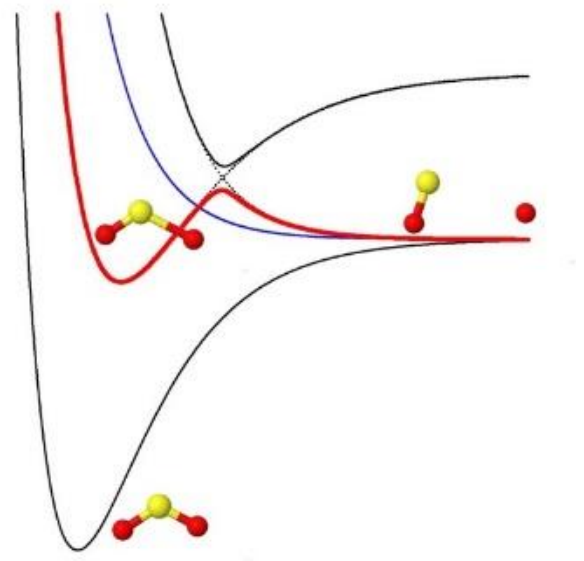